\documentclass[pre,twocolumn]{revtex4}
\usepackage{graphicx}
\usepackage{dcolumn}
\usepackage{bm}
\usepackage{xcolor}
\usepackage{soul}
\usepackage[english]{babel}
\usepackage{epsfig}
\usepackage{amsmath}
\usepackage{amsfonts}
\usepackage{amssymb}
\usepackage{float}
\usepackage{mathrsfs}
\usepackage{mathtools}

\begin{document}

\title{Solitons in a box-shaped wavefield with noise: perturbation theory and statistics}

\author{Rustam Mullyadzhanov$^{1,2}$}\email{rustammul@gmail.com}
\author{Andrey Gelash$^{3,4}$}\email{agelash@gmail.com}

\affiliation{$^{1}$Institute of Thermophysics SB RAS, Novosibirsk 630090, Russia}
\affiliation{$^{2}$Novosibirsk State University, Novosibirsk 630090, Russia}
\affiliation{$^{3}$Institute of Automation and Electrometry SB RAS, Novosibirsk 630090, Russia}
\affiliation{$^{4}$Skolkovo Institute of Science and Technology, Moscow 121205, Russia}

\begin{abstract}
We investigate the fundamental problem of the nonlinear wavefield scattering data corrections in response to a perturbation of initial condition using inverse scattering transform theory.
We present a complete theoretical linear perturbation framework to evaluate first-order corrections of the full set of the scattering data within the integrable one-dimensional focusing nonlinear Schr\"odinger (NLSE) equation.
The general scattering data portrait reveals nonlinear coherent structures – solitons  – playing the key role in the wavefield evolution.
Applying the developed theory to a classic box-shaped wavefield we solve the derived equations analytically for a single Fourier mode acting as a perturbation to the initial condition, thus, leading to the sensitivity closed-form expressions for basic soliton characteristics, i.e. the amplitude, velocity, phase and its position.
With the appropriate statistical averaging we model the soliton noise-induced effects resulting in compact relations for standard deviations of soliton parameters.
Relying on a concept of a virtual soliton eigenvalue we derive the probability of a soliton emergence or the opposite due to noise and illustrate these theoretical predictions with direct numerical simulations of the NLSE evolution.
The presented framework can be generalised to other integrable systems and wavefield patterns.
\end{abstract}

\maketitle

The propagation of nonlinear waves is well-described by a number of integrable models leading to the concept of the scattering data also known as the nonlinear Fourier spectrum.
Inverse scattering transform theory uncovers a trivial evolution of this spectrum and provides an elegant integration method, for example, for the one-dimensional  Korteweg--de  Vries  (KdV) and nonlinear Schr\"odinger (NLSE) equations representing fundamental models of nonlinear physics \cite{NovikovBook1984, AblowitzBook1981}.
The scattering data portrait reveals nonlinear coherent structures -- solitons, which are parametrized by eigenvalues and norming constants as well as dispersive waves described by the reflection coefficient.
Solitons represent the backbone of the evolution of water wave groups \cite{OsborneBook2010, slunyaev2006nonlinear, Randoux2018nonlinear, slunyaev2018analysis, suret2020nonlinear} or propagation of light pulses in a fiber \cite{turitsyn2008soliton, derevyanko2012nonlinear, Randoux2018nonlinear}.
These nondispersive waves play the key role in nonlinear features such as the formation of rogue waves \cite{akhmediev2009excite, OsborneBook2010, Gelash2018} and considered as the main carriers of information in nonlinear optical telecommunication systems \cite{yousefi2014information, Turitsyn2017nonlinear,Frumin2017}.
In practice the wavefield typically evolves in the presence of noise altering the scattering data and leading to the important issue of sensitivity \cite{KivsharRMP1989, derevyanko2016capacity, Aref2019, chekhovskoy2019nonlinear, Turitsyn2020}.
For KdV and NLSE models the perturbation theory has been developed in the case of small continuous pumping or dissipation \cite{KaupSIAM1976, KarpmanJETP1977, keener1977solitons, karpman1979soliton, anderson1978variat, bondeson1979soliton, karpman1981perturbational, Bass1988,KivsharRMP1989,Kaup1990} as well as for an instant perturbation
\cite{KivsharRMP1989, YangBook2010, Aref2019}, see also some recent advancements \cite{nemykin2008influence,grinevich2018finite}.
However, the analytical insight for perturbed scattering data is still missing even for simple model problems.
In this work we develop the perturbation theory for a basic rectangular (box) wavefield initially perturbed by stochastic noise within the focusing NLSE.
The evolution of a box field within the NLSE model representing a classical so-called dam-break problem \cite{whitham2011linear} attracts experimental attention in optics \cite{xu2017dispersive} and hydrodynamics \cite{bonnefoy2020modulational} as well as theoretical efforts \cite{el2016dam}.
A wide box-shaped field is unstable to long wave perturbations constituting the modulation instability \cite{benjamin1967disintegration, zakharov2009modulation} typically stimulated by adding noise \cite{Kraych2019}.
We provide a complete first-order perturbation ansatz for the full scattering data
including soliton parameters: amplitudes, velocities, phases and positions.
The derived equations are solved analytically for a box field perturbed by a single Fourier mode.
Then using statistical averaging we model the effect of noise on solitons resulting in compact expressions for standard deviations.
Finally, using a concept of a virtual soliton eigenvalue we derive the probability of a noise-induced soliton emergence event or disappearance revisiting a fundamental problem using a new tool \cite{Wright1980, Bussac1985, Larroche1986, kaup1986forced, KivsharRMP1989}.
We write the focusing NLSE for a complex wavefield $q(t, x)$ in a non-dimensional form:
\begin{eqnarray}
\label{eqNLS} i q_t + \frac12 q_{xx} + |q|^2 q = 0,
\end{eqnarray}
where $t$ and $x$ are the time and spatial coordinate.
The scattering data can be found with the direct scattering transform (DST) based on the Zakharov--Shabat (ZS) equation \cite{Zakharov1972} representing an auxiliary linear system for a vector wave function $\Phi = (\phi_1,\phi_2)^T$
\begin{eqnarray}
\mathcal{L} \Phi - \zeta \Phi = 0, \;\;\; \mathcal{L} = 
\begin{pmatrix}
i \partial_x   & -i q(x) \\ -i q^*(x)   & - i \partial_x
\end{pmatrix},
\label{ZSsystem}
\end{eqnarray}
where $\zeta = \xi + i \eta$ is the time-independent complex spectral parameter with real $\xi$ and $\eta$, the superscripts $T$ and the star stand for a transposition and complex conjugation.
Eq. (\ref{ZSsystem}) is typically solved for a fixed moment of time $t_0$ with $q(x) = q(t_0, x)$ playing the role of a potential.

In case of potentials with compact support the wave function has the following asymptotics \cite{faddeev2007hamiltonian}:
\begin{equation}
\label{BoundCond}
\Phi |_{x \to -\infty} = (e^{- i \zeta x}, 0)^T,
\,\,\,
\Phi |_{x \to \infty} = (a  e^{-i \zeta x}, b  e^{i \zeta x})^T.
\end{equation}
The scattering coefficients $a(\zeta)$ and $b(\zeta)$ are connected to the scattering data $\{ \zeta_n, \rho_n; r \}$ as follows: 
\begin{equation}
\label{ScatteringData}
a(\zeta_n)=0,
\,\,\,\,
\rho_n = b(\zeta_n) / a'(\zeta_n);
\,\,\,\,
r(\xi) = b(\xi)/ a(\xi),
\end{equation}
where $\{ \zeta_n, \rho_n \}$ is a countable set of eigenvalues (discrete spectrum) and associated norming constants, while $r(\xi)$ is the reflection coefficient defined on a real axis (continuous spectrum).
Each eigenvalue $\zeta_n = \xi_n + i\eta_n$ corresponds to a soliton with the amplitude $2\eta_n$,
group velocity $2\xi_n$ while the position and phase are characterized by $\rho_n$, see \cite{NovikovBook1984}.
The condition $a(\zeta_n)=0$ with $\{ \eta_n \} > 0$ for $n = 1, ..., N$ guaranties the decay of the wave function according to asymptotics (\ref{BoundCond}) leading to physically meaningful soliton eigenvalues $\{ \zeta_n \}$  \cite{NovikovBook1984}.
At the same time, the condition $a(\zeta_n)=0$ can also be satisfied for $\{ \eta_n \} <0$ with $n = -1,-2,...$, see also \cite{tsoy2003interaction}, with the exponentially growing wave function (\ref{BoundCond}).
We refer to these $\{\zeta_n \}$ distinguished by negative indexes $n$ as nonphysical zeros of $a(\zeta)$ or virtual soliton eigenvalues, the number of which can be inifinite.
With the inner product of two vectors $\langle \Psi, \Phi \rangle = \int_{-\infty}^{\infty} \Psi^{* T} \Phi d x$ we derive an eigensystem adjoint to (\ref{ZSsystem}):
\begin{eqnarray}
\mathcal{L}^\dagger \Phi^\dagger - \zeta^* \Phi^\dagger = 0, \;\;\; \mathcal{L}^\dagger = 
\begin{pmatrix}
i \partial_x   & i q \\ i q^*   & - i \partial_x
\end{pmatrix},
\label{ZSadjoint}
\end{eqnarray}
where the adjoint operator satisfies the relation $\langle \Phi^\dagger, \mathcal{L} \Phi \rangle = \langle \mathcal{L}^\dagger \Phi^\dagger, \Phi \rangle$.
Note that $\Phi^\dagger = (\phi_2^*, \phi_1^*)^T$ satisfies Eq. (\ref{ZSadjoint}).
We are interested in the variation of $\{ \zeta_n, \rho_n \}$ and $r$ associated with a small perturbation $\delta q(x)$ of the potential.
Let us take the variation of Eq. (\ref{ZSsystem}):
\begin{eqnarray}
\delta\big( \mathcal{L} \Phi - \zeta \Phi \big) = (\delta \mathcal{L} - \delta \zeta ) \Phi + ( \mathcal{L} - \zeta ) \delta \Phi = 0.
\label{ZSvar}
\end{eqnarray}
To cancel out the second term in the last expression we take the inner product of Eq. (\ref{ZSvar}) with $\Phi^\dagger$ resulting in $\langle \Phi^\dagger, (\delta \mathcal{L} - \delta \zeta ) \Phi \rangle = 0$.
Extracting $\delta \zeta$, we end up with the following expression \cite{KivsharRMP1989}:
\begin{eqnarray}
\delta \zeta = \frac{\langle \Phi^\dagger, \delta \mathcal{L} \Phi \rangle}{\langle \Phi^\dagger, \Phi \rangle},  \;\;\; \delta \mathcal{L} = 
-i \begin{pmatrix}
0 & \delta q \\ \delta q^* & 0
\end{pmatrix},
\label{dzeta}
\end{eqnarray}
where $\langle \Phi^\dagger, \delta \mathcal{L} \Phi \rangle = - i \int_{-\infty}^{\infty} (\phi_1^2 \delta q^* + \phi_2^2 \delta q) d\tau$ and $\langle \Phi^\dagger, \Phi \rangle = \int_{-\infty}^{\infty} 2 \phi_1 \phi_2 d\tau$.
A deviation in $\zeta$ leads to small changes in $\Phi$ and as a consequence to non-zero $\delta a$ and $\delta b$ as well as their derivatives with respect to $\zeta$.
To find the perturbation $\delta \rho$, we take the variation of $\rho$: 
\begin{eqnarray}
\delta \rho = \delta b / a' - b \delta a' / a'^2 = \rho ( \delta b / b - \delta a' / a' ).
\label{drho}
\end{eqnarray}
The values $b$ and $a'$ at some $\{ \zeta_n \}$ are assumed to be already known.
According to boundary conditions, see (\ref{BoundCond}), in order to obtain $\delta b$ and $\delta a'$ we have to explore the variation of the solution $\Phi$ and $\Phi' = \partial_\zeta \Phi$ at $x \to \infty$. 
However, a multiplier $e^{i \zeta x}$ in boundary conditions does not make it straightforward.
We rewrite the ZS system using a new variable \cite{vaibhav2018higher}:
\begin{eqnarray}
\label{Phi2PhiTilde}
\widetilde{\Phi} = e^{i \zeta \Lambda x} \Phi, \;\;\; \text{where} \;\;\; \Lambda = \mathrm{diag}(1,-1),
\label{newPhi}
\end{eqnarray}
leading to the system
\begin{eqnarray}
\partial_x \widetilde{\Phi} = \widetilde{Q}_{-} \widetilde{\Phi}, \;\;\; \widetilde{Q}_{\pm} = 
\begin{pmatrix}
0 & q e^{+} \\ \pm q^* e^{-} & 0
\end{pmatrix},
\label{newZS}
\end{eqnarray}
where we use the notation $e^{\pm} = e^{\pm 2 i \zeta x}$ and $\widetilde{Q}_{\pm}$.
We arrive to a modified set of boundary conditions:
\begin{equation}
\label{newBC}
\widetilde{\Phi} |_{x \to -\infty} = (1,0)^T
,\,\,\,\,\,\,\,\,
\widetilde{\Phi} |_{x \to \infty} = (a,b)^T.
\end{equation}
This important simplification let us express $\delta b$ and $\delta a'$ using variations $\delta \widetilde{\Phi}$ and $\delta \widetilde{\Phi}'$ at $x \to \infty$:
\begin{equation}
\label{newBC2}
\delta \widetilde{\Phi} |_{x \to \infty} = (\delta a, \delta b)^T,
\,\,\,\,\,\,\,
\delta \widetilde{\Phi}' |_{x \to \infty} = (\delta a', \delta b')^T.
\end{equation}
Thus, to obtain $\delta \rho$ we have to compute $\delta \widetilde{\Phi}(x)$ and $\delta \widetilde{\Phi}'(x)$.
Taking the variation of Eq. (\ref{newZS}), we obtain:
\begin{eqnarray}
\partial_x \delta \widetilde{\Phi} = \widetilde{Q}_{-} \delta \widetilde{\Phi} + \delta \widetilde{Q}_{-} \widetilde{\Phi},
\label{eqdPhi}
\end{eqnarray}
arriving to a nonhomogeneous equation for $\delta \widetilde{\Phi}$.
To derive the equation for $\delta \widetilde{\Phi}'$ we first differentiate Eq. (\ref{newZS}) with respect to $\zeta$ and then take the variation since these operations do not commute:
\begin{eqnarray}
\partial_x \delta \widetilde{\Phi}' = \widetilde{Q}_{-} \delta \widetilde{\Phi}' + \delta \widetilde{Q}_{-} \widetilde{\Phi}' + \widetilde{Q}'_{-} \delta \widetilde{\Phi} + \delta \widetilde{Q}'_{-} \widetilde{\Phi}.
\label{eqdPhip}
\end{eqnarray}
According to (\ref{newBC}), at $x \to -\infty$ zero boundary conditions have to be imposed for $\delta \widetilde{\Phi}$ and $\delta \widetilde{\Phi}'$. 
The full expressions for the matrices are as follows:
\begin{eqnarray}
\label{Qmatrices}
&& \widetilde{Q}'_{-} = 2 i x \widetilde{Q}_{+}, \;\;\;  \delta \widetilde{Q}_{-} =
\begin{pmatrix}
0   &   \delta q e^{+}  \\ - \delta q^* e^{-}   &   0
\end{pmatrix} + 2 i \delta \zeta x \widetilde{Q}_{+}, \nonumber\\
&& \delta \widetilde{Q}'_{-} =
2 i x
\begin{pmatrix}
0   &   \delta q e^+ \\ \delta q^* e^-  &  0
\end{pmatrix} - 4 \delta \zeta x^2 \widetilde{Q}_{-}.
\end{eqnarray}
We extend the treatment for Eqs. (\ref{eqdPhi}), (\ref{eqdPhip}) to find $\delta b$ and $\delta a'$ appearing in Eq. (\ref{drho}).
Using both independent solutions of the ZS system \cite{NovikovBook1984}, i.e. $\Phi = (\phi_1, \phi_2)^T$ and 
\begin{eqnarray}
\Psi = (\psi_1, \psi_2)^T = (-\phi^*_2, \phi^*_1)^T|_{\zeta = \zeta^*},
\end{eqnarray}
we represent the solution of Eq. (\ref{eqdPhi}) as 
\begin{eqnarray}
\delta \widetilde{\Phi} = f_1(x) \widetilde{\Phi} + f_2(x) \widetilde{\Psi},
\label{dPhi_f}
\end{eqnarray}
where variables with tilde are obtained in agreement with Eq. (\ref{newPhi}) and $f = (f_1, f_2)$ is to be determined.
Substituting this form of $\delta \widetilde{\Phi}$ to Eq. (\ref{eqdPhi}) and using Eq. (\ref{newZS}) and the notation $\mathcal{W}$, we obtain:
\begin{eqnarray}
f'_1 \widetilde{\Phi} + f'_2 \widetilde{\Psi} = \mathcal{W} f' = \delta \widetilde{Q}_{-} \widetilde{\Phi},
\,\,\,\,\, \mathcal{W} = (\widetilde{\Phi}^T, \widetilde{\Psi}^T).
\label{fEq}
\end{eqnarray}
The solution of Eq. (\ref{fEq}) is as follows:
\begin{eqnarray}
f(x) = \int_{-\infty}^{x} \mathcal{W}^{-1}(y) \delta \widetilde{Q}_{-}(y) \widetilde{\Phi}(y) dy,
\label{fEqsol}
\end{eqnarray}
where the integration constant is zero due to zero boundary conditions of $\delta \widetilde{\Phi}$ at $x \to -\infty$.
Using the expression (\ref{fEqsol}), we recover the solution for $\delta \widetilde{\Phi}$, see (\ref{dPhi_f}).
A similar scheme can be applied to Eq. (\ref{eqdPhip}) with the form $\delta \widetilde{\Phi}' = g_1 \widetilde{\Phi} + g_2 \widetilde{\Psi}$ where for $g = (g_1, g_2)$ we can obtain:
\begin{eqnarray}
g(x) = \int_{-\infty}^{x} \mathcal{W}^{-1} (\delta \widetilde{Q}_{-} \widetilde{\Phi}' + \widetilde{Q}'_{-} \delta \widetilde{\Phi} + \delta \widetilde{Q}'_{-}  \widetilde{\Phi}   ) dy.
\label{gEqsol}
\end{eqnarray}

The perturbation of the reflection coefficient is expressed as follows:
\begin{eqnarray}
\delta r = \delta b / a - b \delta a / a^2 = r ( \delta b / b - \delta a / a ).
\label{drcont}
\end{eqnarray}
Note that $\delta r$ is defined on the real axis $\xi$, thus, $\delta \zeta = 0$ in Eqs. (\ref{eqdPhi}), (\ref{eqdPhip}), see also expressions (\ref{Qmatrices}).
As the basic unperturbed potential we consider a box function $q_\sqcap(x) = A$ for $|x| < L / 2$ where $A$ is a real-valued constant, while $q_\sqcap = 0$ otherwise.
The scattering coefficients are as follows \cite{manakov1973nonlinear}:
\begin{eqnarray}
\label{aBox}
&& a(\zeta) = e^{i L \zeta} \big( \cos(\chi L) - i \zeta\sin(\chi L)/\chi \big), \\
&& b(\zeta) = - A \sin(\chi L)/\chi, \;\;\;\;\; \chi = \sqrt{A^2 + \zeta^2}.
\label{bBox}
\end{eqnarray}
The wave function in the region $|x| < L/2$ represents:
\begin{eqnarray}
\label{solPhiBox}
&& \Phi = e^{\frac{i \zeta L}{2}}\frac{\nu}{1 - \nu^2}
\begin{pmatrix}
\nu^{-1} e^{-i \mu} - \nu e^{i \mu} \\ e^{-i \mu} - e^{i \mu}
\end{pmatrix}, \\
&& \nu = i (\zeta - \chi)/A, \;\;\;\;\;\;\; \mu = \chi L/2 + \chi x,
\end{eqnarray}
while for $|x| > L/2$ it corresponds to the asymptotics (\ref{BoundCond}).

The condition $a(\zeta_n) = 0$ for both cases $\{ \eta_n \} > 0$ and $\{ \eta_n \} < 0$ leads to the transcendental equation:
\begin{eqnarray}
\label{azeros}
&& \tan (\chi_n L) = - i \chi_n/\zeta_n, \;\;\;\;\; \chi_n = \sqrt{A^2 + \zeta_n^2}.
\end{eqnarray}
Using (\ref{aBox})--(\ref{azeros}) we express norming constants (\ref{ScatteringData}) as:
\begin{eqnarray}
\label{rhon}
\rho_n = - i \chi_n^2 e^{- i \zeta_n L} / [ A (1 - i \zeta_n L) ].
\end{eqnarray}
Note that the number of solitons in the box is limited by $N = \text{Integer}[ 1/2 + A L / \pi]$ and all the eigenvalues are alinged on the imaginary axis, i.e. $\{ \zeta_n \} = i \{ \eta_n \}$ for
$n=1,..,N$ 
and the solitons have zero velocities, see \cite{klaus2002purely, tsoy2003interaction, desaix2003eigenvalues}.
\begin{figure} 
\centering
\includegraphics[width=0.49\linewidth]{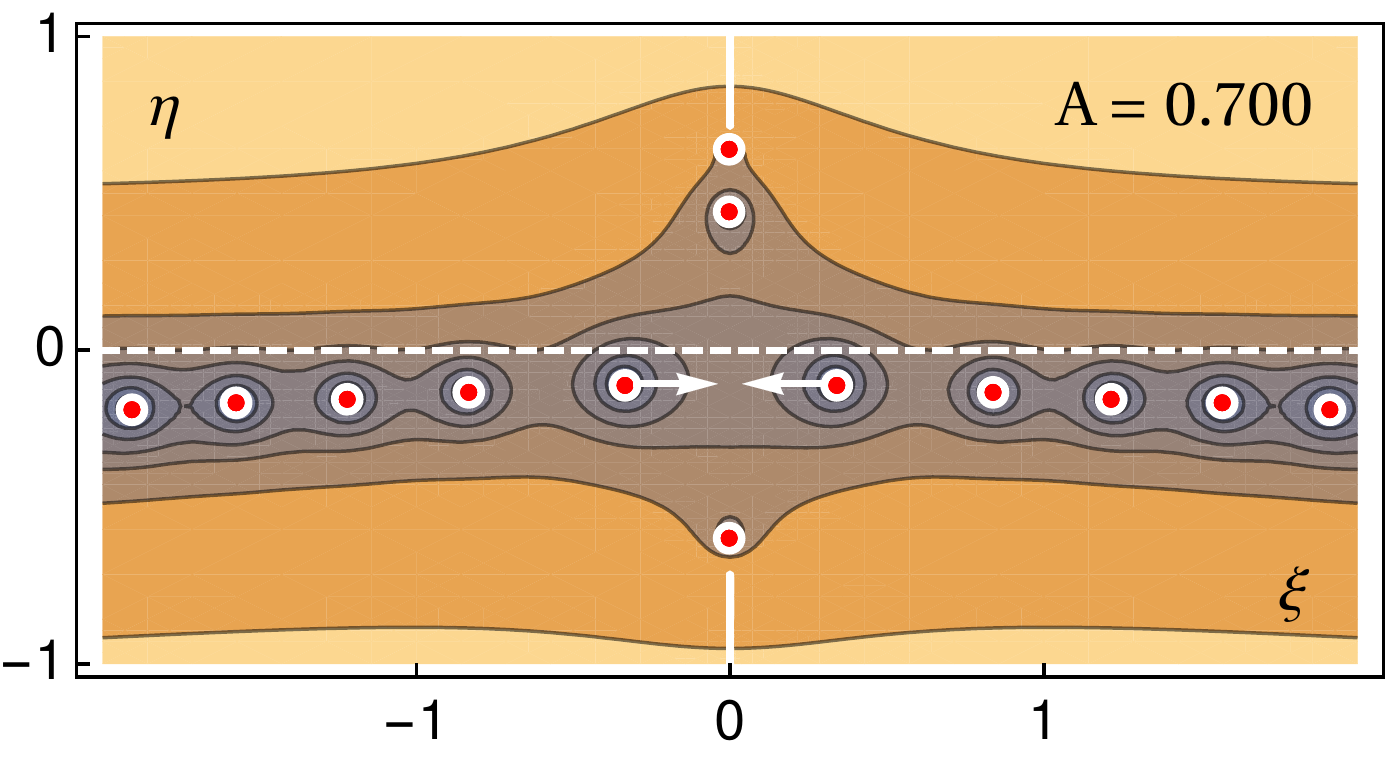}
\includegraphics[width=0.49\linewidth]{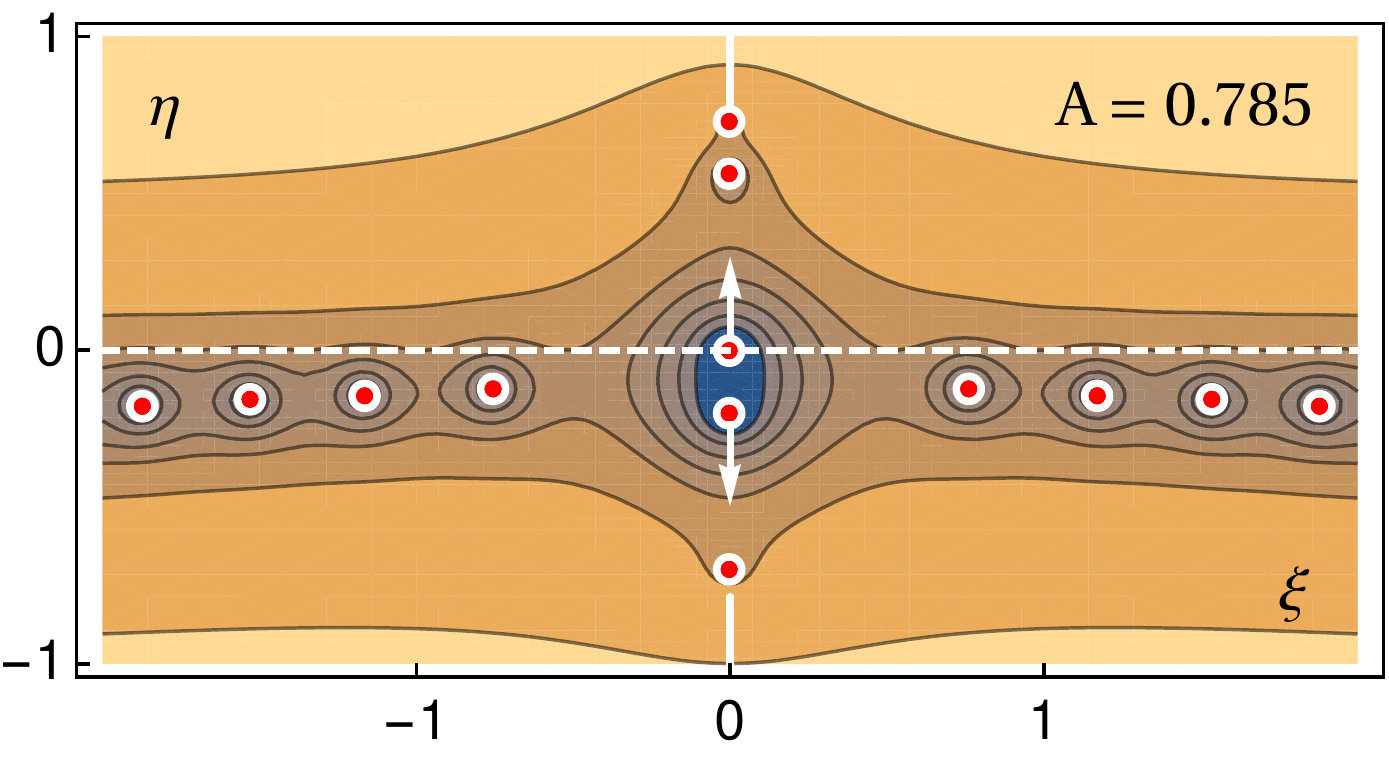}

\includegraphics[width=0.49\linewidth]{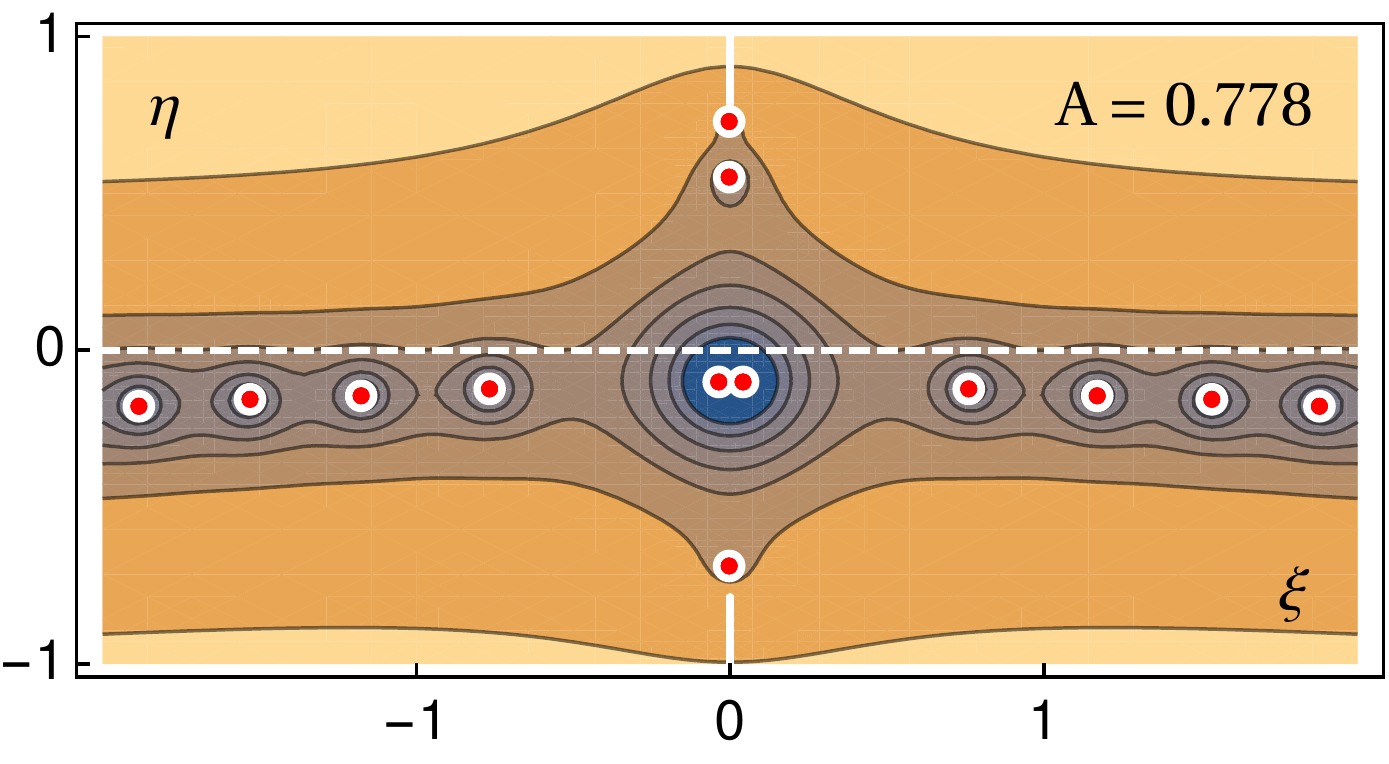}
\includegraphics[width=0.49\linewidth]{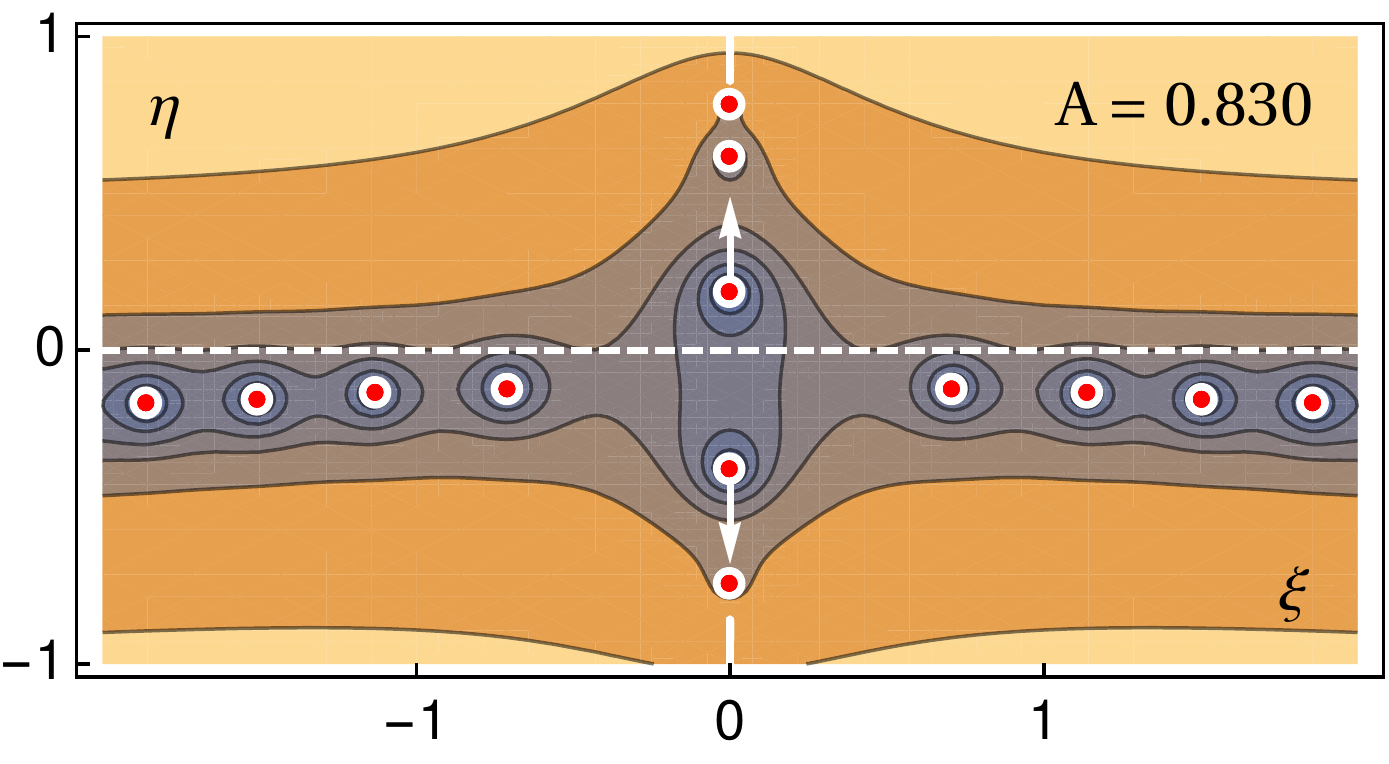}

\includegraphics[width=0.49\linewidth]{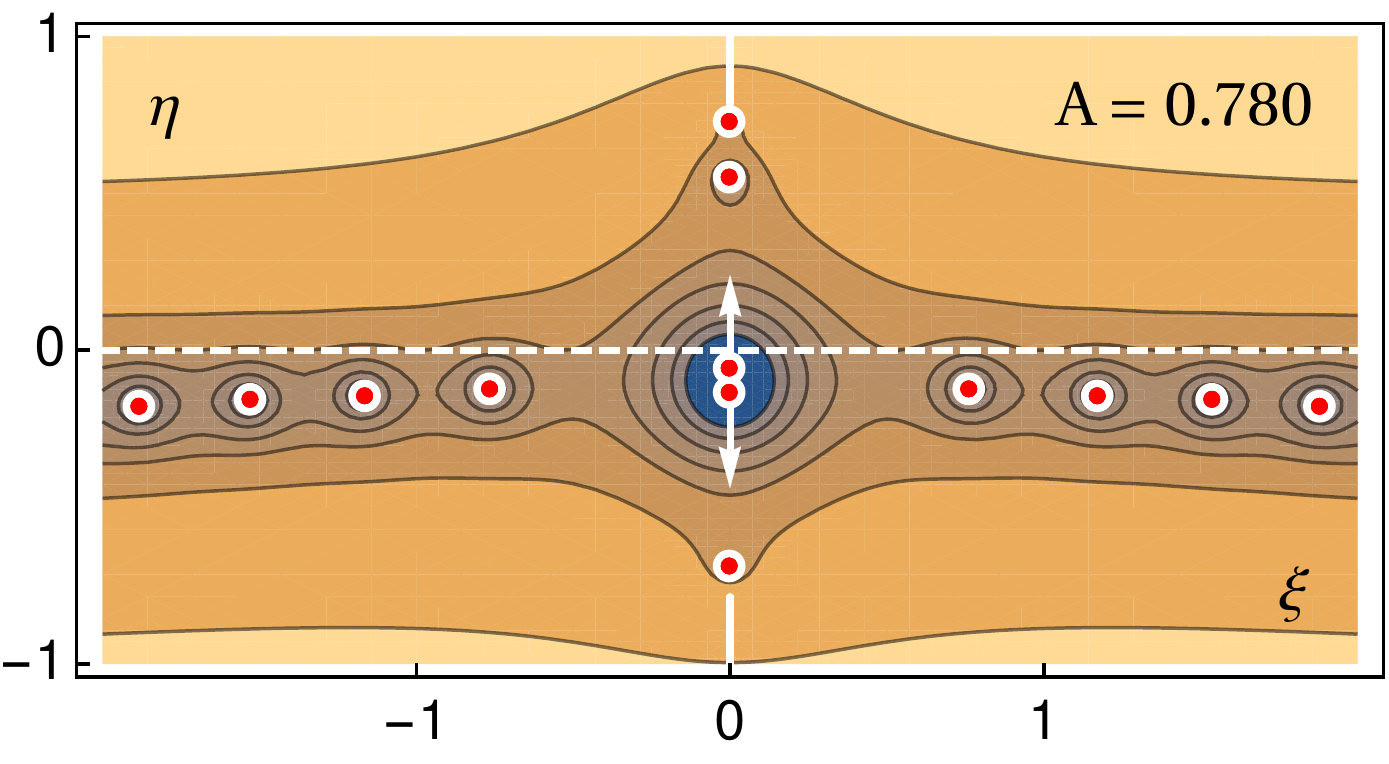}
\includegraphics[width=0.49\linewidth]{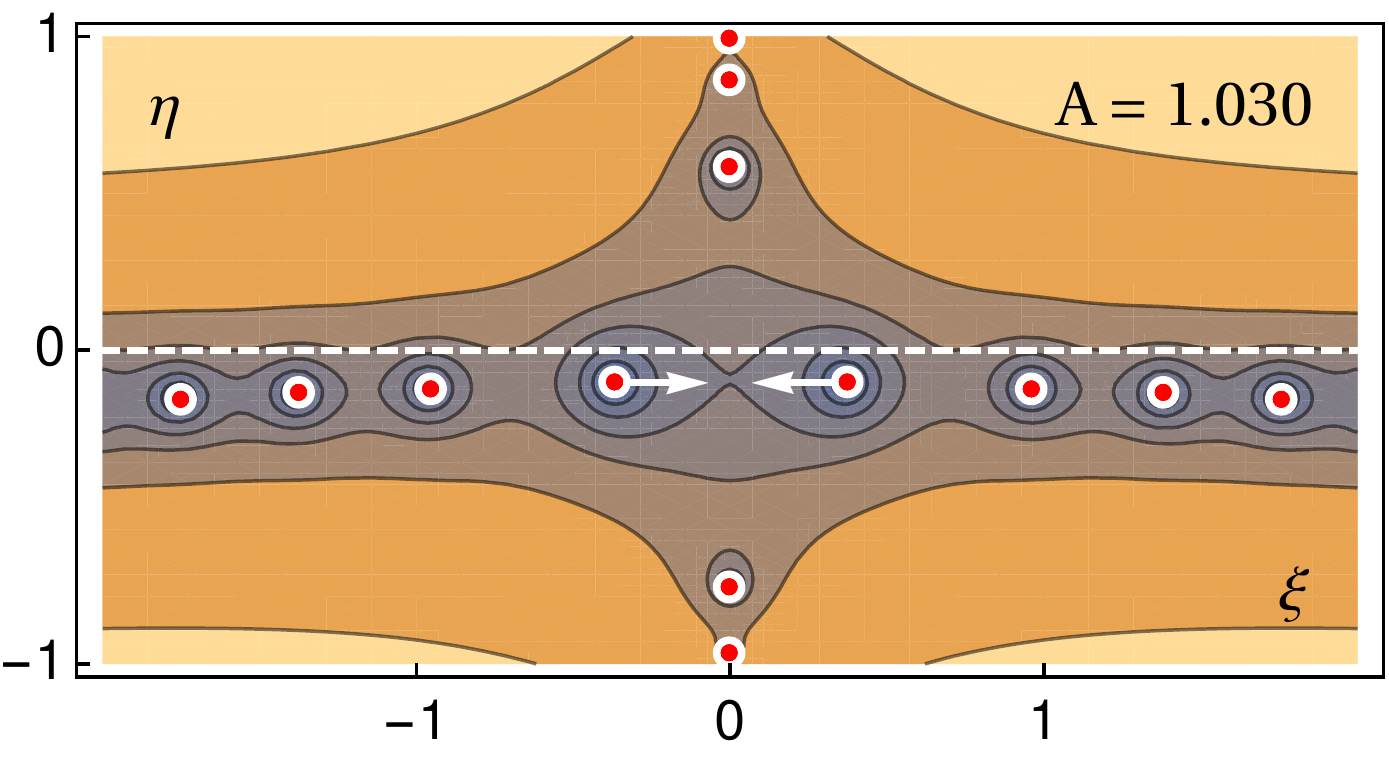}
\caption{
Contours of $| e^{-i L \zeta} a(\zeta) |$ for different values of $A$ and $L = 10$.
Red-in-white points denote the roots of this expression -- physical $\{ \eta_n \} >0$ and virtual $\{ \eta_n \} <0$ soliton eigenvalues.
White dashed line shows the real axis, while white arrows show the direction of a pair of roots in focus with the increase of $A$. 
}
\label{fig_box_eig}
\end{figure}
First we study the evolution of the roots of $e^{-i L \zeta} a(\zeta)$ according to Eq. (\ref{aBox}) for different values of $A$ and $L = 10$, see Fig. \ref{fig_box_eig}.
In addition to physically meaningful zeros with $\{ \eta_n \} >0$ representing solitons, there is a band of roots with $\{ \eta_n \} < 0$ along the real axis and several points on the imaginary axis.
Increasing the value of $A$ from $0.7$ to $0.778$ two symmetric negative zeros approach the imaginary axis and stick together at a slightly higher $A$ forming a degenerate root.
Further they move apart along the imaginary axis, see the case with $A = 0.78$.
At some point a negative zero crosses the real axis and becomes a soliton, see $A = 0.785$ and $0.83$.
At higher $A$ the next pair of negative roots approach the imaginary axis in the lower half of the $\zeta$-plane, see $A = 1.03$, resembling the initial situation with $A = 0.7$.
Fig. \ref{fig_box_eig} illustrates a situation when a simple box-like perturbation moves a nonphysical zero to the region $\eta > 0$ leading to the emergence of a soliton from its virtual counterpart.
As a perturbation we consider real- and imaginary-valued Fourier modes:
\begin{eqnarray}
\label{dqre}
&& \delta q^{\mathrm{re}} = \varepsilon \cos (kx + \varphi), \;\;\;\;\; \delta q^{\mathrm{im}} = i \delta q^{\mathrm{re}},
\label{dqim}
\end{eqnarray}
where $\varepsilon$ is a small parameter while $k$ and $\varphi$ are the wavelength and phase, respectively.
To calculate $\delta \zeta_n$ and $\delta \rho_n$ caused by $\delta q$ as in (\ref{dqre}) we use the explicit form of the wave function (\ref{solPhiBox}) in the relation (\ref{dzeta}) and equations (\ref{eqdPhi}), (\ref{eqdPhip}) with the conditions (\ref{azeros}) employed for algebraic simplifications.
The following exact expressions are obtained:
\begin{eqnarray}
\label{dzetareim}
&& \delta \zeta_{n}^{\mathrm{re}} = i\varepsilon h^{\mathrm{re}}(k,\zeta_n) \cos \varphi, \,\,\, \delta \zeta_{n}^{\mathrm{im}} = \varepsilon  h^{\mathrm{im}}(k,\zeta_n) \sin \varphi, \;\;\;\;\;\;\; \\
\label{drhore}
&& \delta \rho_{n}^{\mathrm{re}}  = i\varepsilon [ s_{1}^{\mathrm{re}}(k,\zeta_n) \cos \varphi + s_{2}^{\mathrm{re}}(k,\zeta_n) \sin \varphi ], \\
\label{drhoim}
&& \delta \rho_{n}^{\mathrm{im}}  = \varepsilon [ s_{1}^{\mathrm{im}}(k,\zeta_n) \cos \varphi + s_{2}^{\mathrm{im}}(k,\zeta_n) \sin \varphi ],
\end{eqnarray}
with the real-valued functions $h^{\mathrm{re/im}}$ and $s_{1,2}^{\mathrm{re/im}}$ given in the Appendix as well as the derivation details, explicit expressions for $\delta r^{\mathrm{re/im}}$ and verification of these results using numerical DST tools \cite{Mullyadzhanov2019, Gelash2020} for the potential $q_\sqcap (1 + \delta q^{\mathrm{re}/\mathrm{im}})$.
Note that according to (\ref{dzetareim})-(\ref{drhoim}), $\delta q^{\mathrm{re}}$ changes only imaginary parts
of $\zeta_n$ and $\rho_n$, while $\delta q^{\mathrm{im}}$ affects on their real pars.
The formulas (\ref{dzetareim})-(\ref{drhoim}) are valid for both physical and virtual soliton eigenvalues.
In the latter case they describe the migration of nonphysical zeros which might result in a birth of a new soliton, similar to the situation illustrated in Fig. \ref{fig_box_eig}.

\begin{figure} 
\centering
\includegraphics[width=0.49\linewidth]{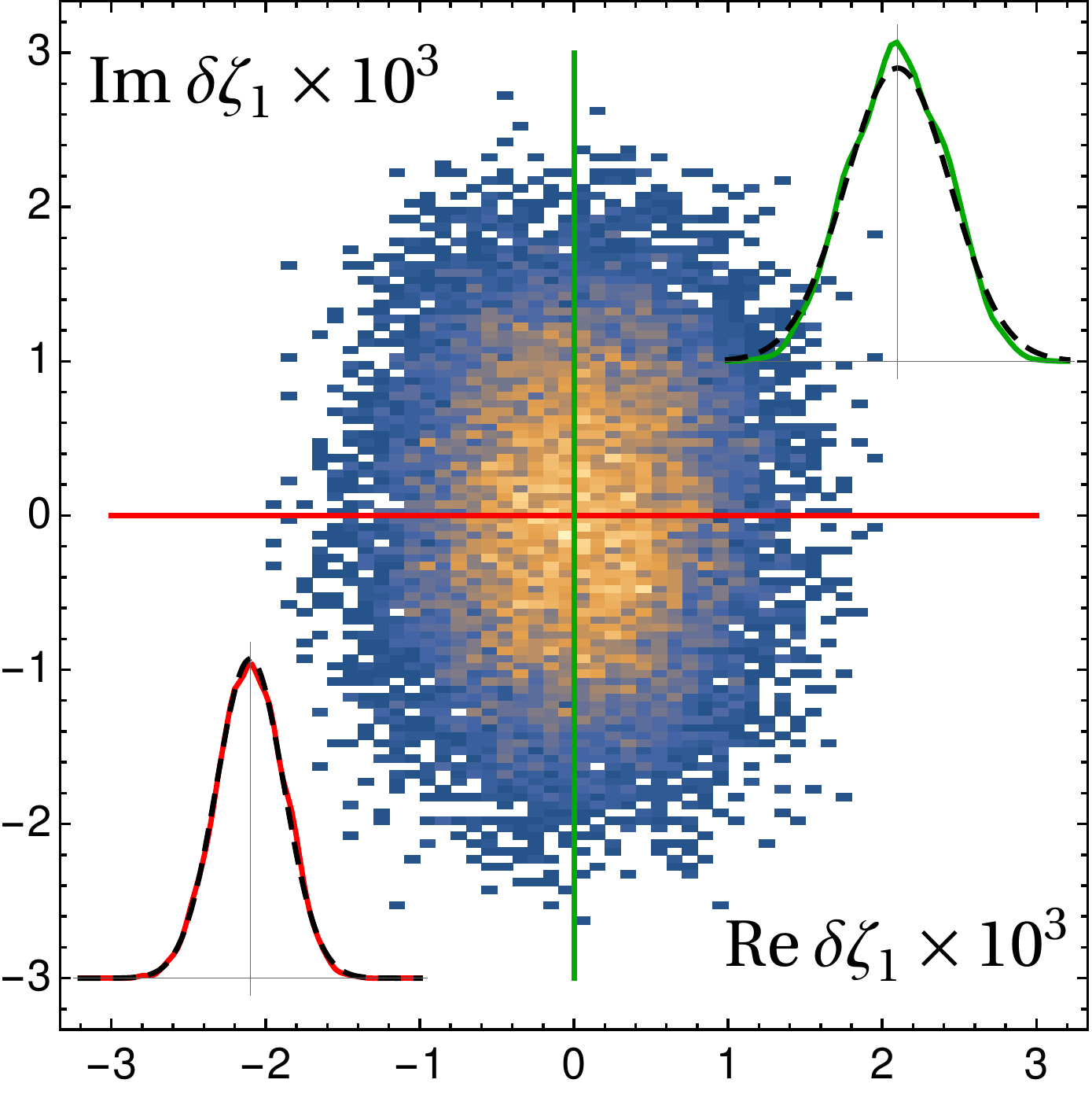}
\includegraphics[width=0.49\linewidth]{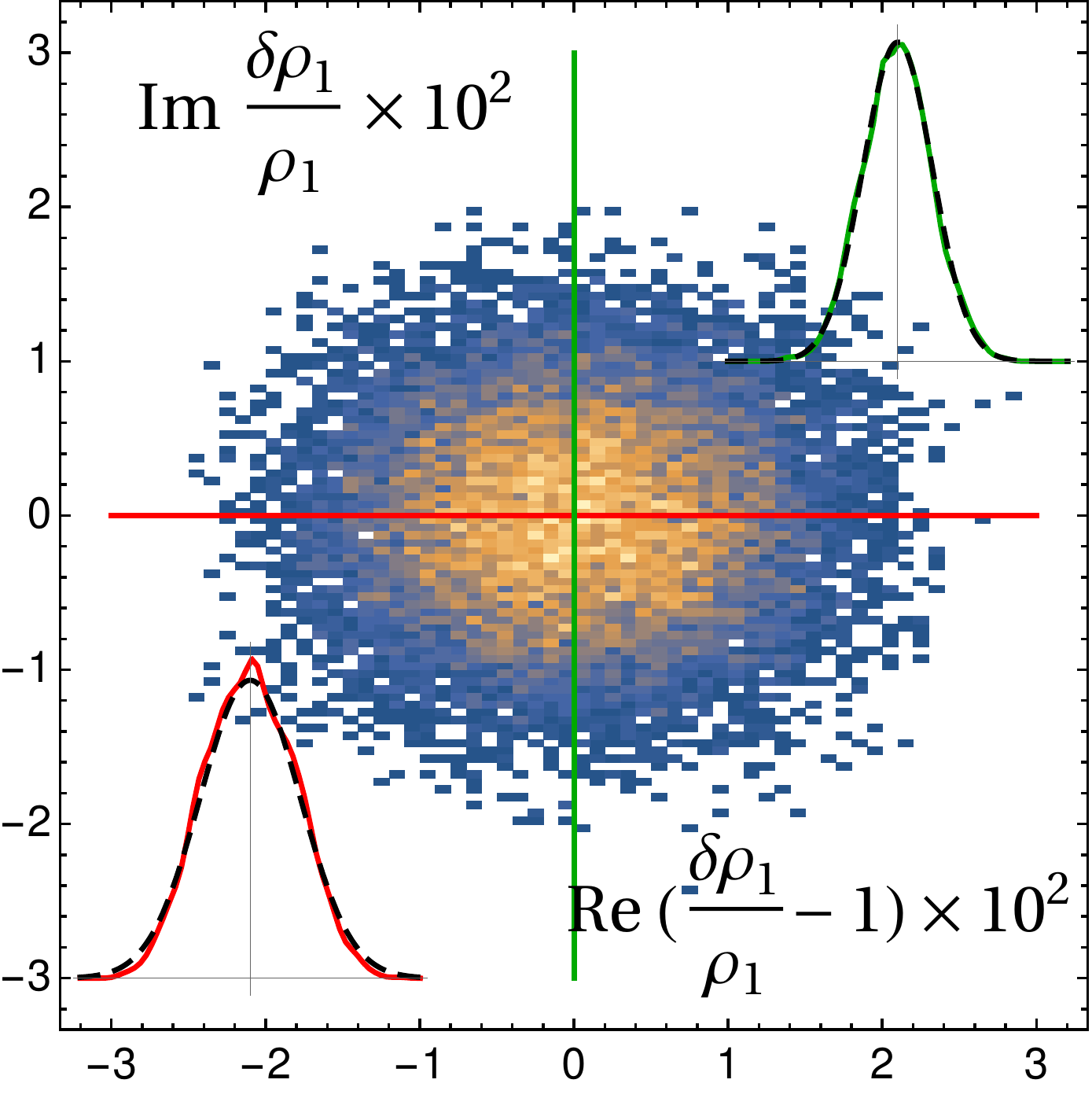}
\caption{
Soliton scattering data deviations induced by $10^4$ realizations of noise superimposed on the box with $A=1$ and $L=12$ computed numerically.
The insets show numerical (green  solid) and  theoretical (grey  dashed) PDFs for soliton parameters.
The unperturbed values of $\zeta_1$ and $\rho_1$ are computed using (\ref{azeros}) and (\ref{rhon}).
}
\label{fig_clouds}
\end{figure}
We consider a sum of modes (\ref{dqre}) with random phases and distributed as $\mathcal{F}(k)$ with respect to $k$.
Integrating (\ref{dzetareim})-(\ref{drhoim}) over $\varphi$, we obtain the following expressions for standard deviations:
\begin{eqnarray}
\label{sigma_zeta}
\label{int_z}
&& (\sigma_{\zeta, n}^{\mathrm{re/im}})^2 = \frac{\varepsilon^2}{2}\int_{-\infty}^{\infty} \mathcal{F} (k) |h^{\mathrm{re/im}} |^2 dk , \\
\label{int_r}
&& (\sigma_{\rho, n}^\mathrm{re/im})^2 = \frac{\varepsilon^2}{2}\int_{-\infty}^{\infty} \mathcal{F} (k) \Big( |s_{1}^\mathrm{re/im} |^2 + |s_{2}^\mathrm{re/im} |^2 \Big) dk. \;\;\;\;\;\;\;\;
\label{sigma_rho}
\end{eqnarray}
These expressions describe the effect of noise on the discrete spectrum for the box potential.
The convergence of (\ref{int_z}), (\ref{int_r}) is guaranteed by an algebraic decay of $\delta \zeta_n$ and  $\delta \rho_n$ for large $k$.
These integrals were evaluated analytically for the white noise model, i.e. $\mathcal{F}(k) = 1$.
The result for $\sigma_{\zeta, n}$ has a compact form (see Appendix for details):
\begin{eqnarray}
\label{sigma_1}
&& (\sigma_{\zeta, n}^{\mathrm{re}})^2 = \frac{\pi \varepsilon^2 \chi^2_n (\eta_n + 2 L A^2)}{2 A^2 (1 + L \eta_n)^2} + \frac{3 \pi \varepsilon^2 \eta_n}{2 (1 + L \eta_n)},
\\
\label{sigma_2}
&& (\sigma_{\zeta, n}^{\mathrm{im}})^2 = \pi \varepsilon^2 \chi^2_n (\eta_n + L A^2) / [ 2 A (1 + L \eta_n)^2 ],
\end{eqnarray}
while $\sigma_{\rho, n}$ is rather cumbersome and omitted in the text.
For a direct comparison with analytical results for $\sigma_{\zeta, n}$ and $\sigma_{\rho, n}$ we simulated a white noise signal as the following normalized collection of $M$ modes with random phases $\varphi_m^{\mathrm{ns},j}$:
\begin{eqnarray}
\label{numerical_noise}
\delta q_{\mathrm{ns}, j} (x) = \varepsilon \sqrt{\Delta k} \sum_{m=1}^{M}
\cos(x \Delta k m + \varphi^{\mathrm{ns}, j}_m ),
\end{eqnarray}
where the subscript `$\mathrm{ns},j$' denotes a particular $j$th set of random phases.
For each case of $10^4$ realizations of the complex-valued noise $\delta q(x) = \delta q_{\mathrm{ns},1}(1 + i \delta q_{\mathrm{ns},2})$ with $\varepsilon=0.005$, $\Delta k=0.1$ and $M = 200$ superimposed on top of the box potential with $L=12$, $A=1$, we computed eigenvalues and norming constant using both the developed perturbation theory and numerical DST, see Appendix.
Fig. \ref{fig_clouds} shows statistical results for the scattering data for the first (largest) soliton out of $N = 4$.
The Gaussian probability density functions (PDFs) with theoretical standard deviations (\ref{int_r})-(\ref{sigma_2}) accurately describe the corresponding numerical data.
\begin{figure} 
\centering
\includegraphics[width=0.98\linewidth]{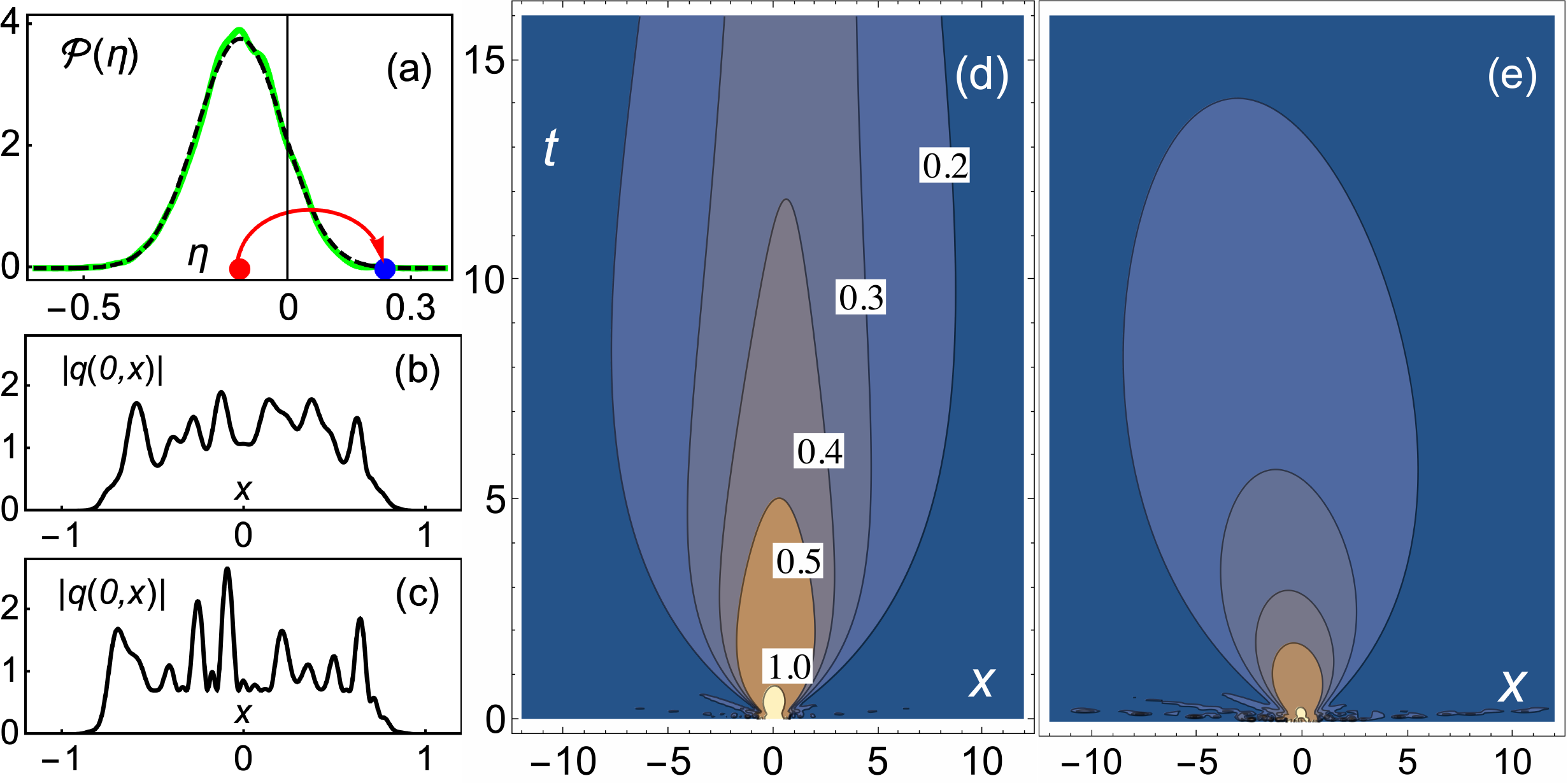}
\caption{
(a) The shift of the nonphysical root to the upper $\zeta$-plane for a particular realization of a real-valued noise within a numerical (green solid) and theoretical (grey dashed) amplitude PDF.
Contour plots of $|q(t,x)|$ from numerical simulations of NLSE.
Evolution of $q_\sqcap$ perturbed by real (b, d) and imaginary (c, e) noise when a soliton is induced and not, respectively.
The theoretically predicted and numerically obtained parameters of the induced soliton are $\zeta_{\mathrm{pr}}=0.223i$, $\zeta_{\mathrm{num}}=0.232i$ with $\rho_{\mathrm{pr}}=-1.275i$, $\zeta_{\mathrm{act}}=-1.311i$.
}
\label{fig_evolution}
\end{figure}
Our theory applied to the nonphysical zeros predicts a birth of a noise-induced soliton.
As an example we consider a box potential with $L=1.46$ and $A=1$ with no solitons and the largest zero $\zeta_{-1} = -0.12i$.
We used $10^4$ realizations of a real-valued noise (\ref{numerical_noise}) which affects only imaginary part of the virtual eigenvalue with $\varepsilon=0.063$, $\Delta k=0.1$ and $M = 500$.
Fig. \ref{fig_evolution}(a) shows theoretical and numerical PDFs for the noise-induced values of $\eta_{-1}$ with the tail $\eta>0$ describing the probability of the emerging soliton with a certain amplitude and norming constant.
We choose a particular noise realization $\delta q_{\mathrm{ns,3}}$, see Fig. \ref{fig_evolution}(b), which shifts the nonphysical zero (green dot) to the upper $\zeta$-plane (red dot) and compute its temporal evolution numerically using NLSE (\ref{eqNLS}) and a standard Runge--Kutta method (see Appendix) with the initial condition $q(0,x) = q_\sqcap(1 + \delta q_{\mathrm{ns,3}})$.
A second computation is performed for the evolution of the initial condition $q(0,x) = q_\sqcap (1 + i \delta q_{\mathrm{ns,3}})$,
shown in Fig. \ref{fig_evolution}(c).
Note that we slightly smoothed $q_\sqcap$ on the edges for numerical simulations, see Appendix for details.
Figs. \ref{fig_evolution}(d,e) show the spatio-temporal contour plots of $|q(t, x)|$ revealing the presence of a strong soliton in the first case, see Fig. \ref{fig_evolution}(d) for parallel contour levels, while in the second case the contours indicate simple decay of the continuous spectrum potential as expected \cite{NovikovBook1984}.
Similarly one can describe migration of the physical root to the nonphysical region, i.e. soliton disappearance.
In this work we presented a complete theoretical framework to evaluate first-order corrections of the full set of scattering data within the NLSE model and applied it to a classic box potential, which can be generalised to other integrable systems and wavefields.
In addition to the classical result for eigenvalues as in (\ref{dzeta}), we derived general expressions, see Eqs. (\ref{eqdPhi}), (\ref{eqdPhip}), leading to the knowledge of soliton phase and position sensitivity.
Starting from a single Fourier mode we obtained statistical integrals (\ref{sigma_zeta}), (\ref{sigma_rho}) allowing to determine the impact of a random-phase noise on soliton parameters which is important in the studies of the spontaneous modulation instability \cite{Agafontsev2015,soto2016integrable,gelash2019bound} and in a number of applications.
The introduction of a concept of a virtual soliton with nonphysical zeros of $a(\zeta)$ allowed us to accurately predict the noise-induced emergence of a soliton.
A similar concept to describe soliton emergence can be further developed for the NLSE model with external pumping, see \cite{bussac1985soliton, kaup1986forced, agafontsev2020growing}.

{\it Acknowledgments} --
First part of the work was supported by Russian Science Foundation grant No. 19-79-10225 (RM for derivation of the perturbation framework).
Second part was supported by Russian Science Foundation grant No. 20-71-00022 (AG for the work on the noise-induced effects).
The authors thank Dr D. Agafontsev for fruitful discussions on virtual soliton eigenvalues.
Statistical simulations were performed at the Novosibirsk Supercomputer Center (NSU).
%


%

\end{document}